\documentclass[preprint,12pt]{aastex}
\usepackage{epsfig}
\usepackage{emulateapj5,apjfonts}
\begin{document}

\def\wdf{white dwarf}
\def\etal{et al.} 
\def\rd{Di\thinspace Stefano}
\newcommand{\chandra}{{\it Chandra}}
\newcommand{\asca}{{\it ASCA}}
\newcommand{\rosat}{{\it ROSAT}}
\newcommand{\sax}{{\it BeppoSAX}}
\newcommand{\xmm}{{\it XMM-Newton}}
\newcommand{\einstein}{{\it Einstein}}
\newcommand{\lum}{\thinspace\hbox{$\hbox{erg}\thinspace\hbox{s}^{-1}$}}
\newcommand{\flux}{\thinspace\hbox{$\hbox{erg}\thinspace\hbox{cm}^{-2}\thinspace\hbox{s}^{-1}$}}

\def\spose#1{\hbox to 0pt{#1\hss}}
\def\laeq{\mathrel{\spose{\lower 3pt\hbox{$\mathchar"218$}}
     \raise 2.0pt\hbox{$\mathchar"13C$}}}
\def\gaeq{\mathrel{\spose{\lower 3pt\hbox{$\mathchar"218$}}
     \raise 2.0pt\hbox{$\mathchar"13E$}}}

\title{Selecting Very Soft X-Ray Sources in External Galaxies: 
Luminous Supersoft X-ray Sources and Quasisoft Sources}

\author{R.~Di\,Stefano$^{1,2}$, A.K.H. Kong$^{1}$} 
\affil{$^{1}$Harvard-Smithsonian Center for Astrophysics, 60
Garden Street, Cambridge, MA 02138}
\affil{$^{2}$Department of Physics and Astronomy, Tufts
University, Medford, MA 02155}

\begin{abstract}

We introduce a procedure to identify very soft X-ray sources (VSSs)
in external galaxies. Our immediate goal was to formulate a
systematic procedure to identify luminous supersoft X-ray sources
(SSSs),   
so as to allow comparisons among galaxies and to study environmental effects. 
The focus of this paper is on the design of the selection algorithm
and on its application to simulated data. In the companion paper we test
it by applying it to sources discovered through {\it Chandra}
observations of $4$ galaxies. 
We find that, in its 
application to both simulated and real data, 
our procedure also selects somewhat harder sources, which we call quasisoft.
Whereas values of $k\, T$ for SSSs are typically tens of eV, some
quasisoft sources (QSSs) may have $k\, T$ 
as high as $\sim 250-300$ eV. 
The dominant spectral component of other 
QSSs may be as soft as SSS spectra, but the spectra
may also include a low-luminosity harder component. We sketch physical models
for both supersoft and quasisoft sources. 
Some SSSs are likely to be accreting white dwarfs; some of these
may be progenitors of Type Ia supernovae. Most QSSs
may be too hot to be white dwarfs. They, together with a
subset of SSSs, may be neutron stars or, perhaps most likely,
accreting intermediate-mass black holes.   

\end{abstract}


\section{Introduction}

\subsection{Goals}

First identified as a class of X-ray sources approximately 11 years
ago, luminous supersoft X-ray sources (SSSs) were characterized by 
$k\, T$ on the order of tens of eV, and $L$ in the range
 $10^{37}-10^{38}$ erg s$^{-1}$.\footnote{See e.g.,
RX J0019.8+2156: Reinsch, Beuermann, \& Thomas 1993, Beuermann et al. 1995; 
RX J0925.7-4758: Motch, Hasinger, \& Pietsch 1994; GQ Mus: Ogelman et
 al. 1993; 1E 1339.8+2837: Hertz, Grindlay, \& Bailyn 1993; AG
 Dra: Greiner et al. 1996; RR Tel: Jordan, Mürset, \& Werner 1994;
 1E 0035.4-7230: Seward \& Mitchell 1981; RX J0048.4-7332: Kahabka,
 Pietsch, \& Hasinger 1994; RX J0058.6-7146: Kahabka et al. 1994; 1E
 0056.8-7154: Wang 1991; RX J0513.9-6951: Schaeidt, Hasinger, \&
 Trümper 1993; CAL 83: Long, Helfand, \& Grabelsky 1981; CAL 87: Long
 et al. 1981; RX J0550.0-7151: Cowley et al. 1993.} 
Because radiation from SSSs is readily absorbed by the interstellar
medium (ISM), we can study large numbers of them only by searching for
them in external galaxies. The spatial resolution of \chandra, and of
\xmm\ as well, make such studies possible. The great distances to
other large galaxies means, however, that the count rates of most of the SSSs
we can detect in them are low. 
Spectral fits are therefore possible for only a small fraction.
The primary goal of the work described in
this paper is to provide a 
well-defined and systematic method to select, from a set of
X-ray sources dominated by low-count sources, 
good SSS candidates. In its present form, the procedure works well  
for both {\it Chandra} and {\it XMM} data. 

\subsection{SSSs in Galaxies}

Soft X-ray sources, including the flagship SSS binaries 
CAL 83 and CAL 87, had been discovered during a survey
of the Magellanic Clouds conducted with the {\it Einstein}
X-ray observatory (Long, Helfand, \& Grabelsky 1981; Seward \& Mitchell 1981).   
The large-scale study of SSSs in other galaxies began with {\it ROSAT}
observations of $\sim 30$ SSSs in the Magellanic Clouds, and in M31. 
Because M31 is similar in many ways to other large
spiral galaxies, it may be a good guide to the sizes of
galactic populations of SSSs.
{\it ROSAT} detected $\sim 15$ SSSs
with luminosities greater than $\sim 10^{37}$ erg s$^{-1}$  in M31\footnote{Subsequent reanalysis
of the {\it ROSAT} data identified additional SSS candidates,
bringing the number to $34$ (Kahabka 1999)}. 
These observations, combined 
with simple models for galactic gas distributions, led to   
an estimate that, at any given time, M31 houses $\sim 1000$ active SSSs
(\rd\ \& Rappaport 1994). 
A similar analysis of the sources and gas distribution  of the Milky Way
found that its  population of SSSs could be comparably large.

The subsequent discovery of the low-luminosity
extension ($10^{35}-10^{36}$ ergs s$^{-1}$) of the class 
(Greiner \etal 1999, Greiner \& \rd 1999, Patterson et al.\, 1998)  may
imply that the Milky Way's 
population of SSSs is at least an order of magnitude 
larger. Recent {\it Chandra} studies (\rd\ 2002, 2003) find convincing
evidence of a lower-luminosity 
component of the class of SSSs in M31 as well.  
If these estimates are correct, and if our Galaxy and M31 are typical,
SSSs may form the single largest galactic population of 
luminous ($L_X > 10^{35}$ erg s$^{-1}$) X-ray  
binaries. 
Whatever their fundamental natures, the likely large sizes of SSSs
populations indicates that they may be pumping 
$10^{40}-10^{41}$ ergs s$^{-1}$ of highly ionizing radiation
into the ISM of their host galaxies. 

\subsection{From Supersoft Sources to Quasisoft Sources}

In the absence of a single definitive
model for SSSs, and with only a small number of known SSSs
in the Galaxy and Magellanic Clouds,  
we did not want to eliminate sources that were somewhat harder
than the canonical sources such as CAL 83.
For reasons summarized in \S 2 and in \S 3.1, we therefore
chose a cut-off in $k\, T$ of $175$ eV for SSSs. We found,
however, that the methods we devised to select sources up to
any given temperature, inevitably also included sources that
either have  (1) somewhat
higher temperatures, but generally in the range $175-250$ eV,
or (2) SSS-like spectra, but with some harder emission as well.   

This suggests two possible results for the method's application to
real data. First, if there are few or no soft sources
just slightly harder than
the canonical SSSs, the procedure will select only SSSs.
Second, if there is a supply of somewhat harder sources, we
will select them as well. 
We have now had opportunities to apply the selection criteria
to the $4$ galaxies studied in the companion paper (\rd\ \& Kong 2003),
to M104 (\rd\ et al. 2003a), M31 (\rd\ et al. 2003b), and roughly
one dozen additional galaxies (\rd\ et al. 2003c). We have found that 
most galaxies
have significant populations of both SSSs and sources with
somewhat harder spectra (e.g., $k\, T < 250$ eV). We refer to the 
latter as quasisoft sources (QSSs).

Our selection procedure distinguishes between SSSs and QSSs
according to which step in the algorithm identifies the source
as being very soft. 
In the galaxies we have studied, 
spectral fits for the brightest SSS and
QSS candidates have verified 
that the algorithmic classification works. 
To simplify the terminology, we
will sometimes use the term ``very soft source" (VSS) to
refer to both SSSs and QSSs.

\subsubsection{Quasisoft Sources}

The physical significance of this new class is not
yet understood, but there are likely to  be several physical models
corresponding to QSSs. First, hot SSSs located behind large gas columns
will have photons in the medium energy band, $M$ ($1.1-2$ keV), 
but may have few
photons in the soft band, $S$ ($0.1-1.1$ keV). For such sources,
the hardness ratios
typically used to identify SSSs will have values not normally associated with 
SSSs, even though their intrinsic characteristics clearly
place them in the SSS category.
Second if 
the detector has poorer than anticipated sensitivity to soft photons,
soft sources can appear to be harder than they actually are.
Thus, some QSSs are likely to have the same physical characteristics
as some other sources identified as SSSs. 
Finally, some QSSs are likely to be genuinely harder than
SSSs, so hard that white dwarf models can be ruled out.
As we will discuss in \S 2, intermediate mass black hole models may
be appropriate for such systems, but neutron star or stellar mass
black hole models should also be considered.
 
\subsection{Open Questions} 

Below we list 
some of the questions we hope
to answer with studies that compare VSS 
populations in different galaxies.

\noindent (1) What are typical galactic populations of SSSs and QSSs?
Irrespective of their fundamental natures,
the answer to this question will allow us to estimate 
the influence of soft X-ray sources as 
ionizers of the ISM.

\noindent (2) 
Are any spiral galaxy parameters 
related to the relative sizes of SSS  and QSS populations?
Answering this question can provide insight
into the age of the populations that spawn very soft sources, and hence might 
help to illuminate their nature. 
\rd\ \& Rappaport (1994) suggested that for spiral
galaxies, the size of
the SSS population might scale with blue luminosity, but this 
has not been tested.  

\noindent (3) Do elliptical galaxies house large SSS/QSS populations?  
Although it has been suggested
that the diffuse soft emission in ellipticals may be due
to SSSs (see, e.g., Fabbiano, Kim, \& 
 Trinchieri 1994), we still know very little about SSSs in ellipticals.
If accreting WDs form the largest segment of SSS populations,
and if a significant fraction of the donor stars have masses
small enough to be typical of the stars found in elliptical galaxies,  
then we may expect SSSs to be important parts of the X-ray 
source population in ellipticals. 

\noindent (4) Within spiral galaxies, what are the relative populations 
of SSSs/QSSs in the galaxy bulges and disks?  

\noindent (5) Do galaxies with massive central black holes
have more SSSs or QSSs located within 1 kpc of the nucleus than
comparable galaxies without massive central black holes?
It has been suggested
that some SSSs within the central kpc  of galaxies which harbor
massive black holes may actually be
the stripped cores of stars that have been tidally disrupted
(\rd\ et al. 2001).
Verification of this hypothesis by studying individual SSSs
will be difficult, so statistical studies
of SSS populations in a large number of galaxies may  provide
the best tests.    

\noindent (6) For all galaxies, are the positions of QSSs and SSSs
correlated to the positions of other objects,
such as HII regions, planetary nebulae, supernova
remnants, or globular clusters? The distances to
 most external galaxies
are too large to allow for convincing
optical identifications. It is nevertheless useful
to identify the types of populations which tend to
be associated with VSSs.
This can provide clues to their fundamental natures.

\noindent (7) Are SSSs significant contributors to the rates of
Type Ia SNe? 
The answer to this question can be achieved by combining information
about typical total galactic populations with studies of the 
viability of the accreting WD models.

\subsection{Previous Work} 

Previous studies of SSSs in external galaxies have used a variety of selection
criteria. In NGC 4697, e.g., Sarazin, Irwin, \& Bregman (2001) 
identified $3$ SSSs by 
requiring that ${\tilde {HR1}}=(\tilde M-\tilde S)/(\tilde M+\tilde S) = -1$ 
and ${\tilde {HR2}}=(\tilde H-\tilde S)/(\tilde H+\tilde S) = -1,$
where $\tilde S,$ $\tilde M,$ and 
$\tilde H$ represent the numbers of counts in the bands
$0.3-1$ keV, $1-2$ keV, and $2-10$ keV, respectively.\footnote{We use tildes in these
expressions because our own definitions of the soft, medium, and hard X-ray bands 
is slightly different (see \S 3).}       
In their studies of the colors of X-ray sources, Prestwich et al.\, (2002),
used the same criteria, which are satisfied by only a handful of
the sources they analyzed, drawn from both M101 and M83. 
Less restrictive criteria were used by Swartz et al.\, (2002) to
identify SSSs in M81. The criteria ${\tilde {HR1}} < -0.5,$ ${\tilde {HR2}} < -0.5$
selected $12$ M81 X-ray sources, $2$ of which were eliminated because they
are identified with foreground stars, while one is identified with
a supernova remnant (SNR). 
Pence et al.\, (2002a) identified $10$ SSSs in M101, but did not
specify the selection criteria. The possible physical interpretation of the
sources seemed to play a role, as one of the galaxy's softest sources
was not counted among the SSSs, perhaps
 because it appears  to be too luminous 
to be a nuclear-burning WD (Pence et al.\, 2002b).
Kong et al.\, (2002a) took another approach,
requiring  
$({\tilde {HR2}} + \sigma_{{\tilde {HR2}}} \leq -1$
{\it and} ($[{\tilde {HR1}} < 0,$] {\it or} [(${\tilde {HR1}} + \sigma_{{\tilde {HR1}}} \leq -0.8]$).
Fourteen sources in the central $17' \times 17'$ of M31 satisfied 
these conditions, of which $2$ were apparently identified with SNRs (Kong et al. 2002b) and $3$
with possible foreground stars.   

These latter criteria were developed in parallel with a study of the entire
SSS population  of M31 as viewed by 
{\it Chandra} (Di~Stefano et al. 2002b, 2003).
It was clear, however, that not all of the sources identified by 
these criteria were equally good candidates for the class.
We therefore began by developing strict selection conditions 
(\rd\ \& Kong 2003 a). These criteria, the so-called HR (hardness ratio)
criteria, are the first set of conditions applied in the algorithm
presented here.   
The full algorithm we present here
allows us to identify SSSs that are, e.g., 
at the high-T end (usually taken to be $\sim 100$ eV) of the class,
even if they are highly absorbed. It also covers cases in which
a small fraction of the emitted radiation is reprocessed and reemitted
in the form of photons of higher energy. In addition the full
algorithm  allows us
to identify sources that may be somewhat hotter than ``classical" SSSs.
This is the first paper to present the full algorithm, and the companion
paper tests its efficacy by studying the results of applying it
to data from $4$ galaxies (\rd\ \& Kong 2003 b).
It has also been applied to {\it Chandra} data from M31 (\rd\ et al. 2003 a),
and from M104 (\rd\ et al. 2003 b).

In \S 2 we give an overview of physical models for VSSs, the class of source
that motivated this work.
In \S 3 we provide a phenomenological definition of SSSs designed to help
select sources with the physical characteristics discussed in \S 2. 
In \S 4 we outline and test a sequential set of selection 
criteria that can be applied
algorithmically to select SSSs as well as sources that are
slightly harder (QSSs), in external galaxies.
\S 5 is devoted to summarizing our conclusions. 
  
\section{Physical Models for Luminous Supersoft X-Ray Sources}

For most classes of astronomical objects, the distinctive phenomenological
characteristics used to identify their members are clearly related to the
distinctive physical characteristics that define their fundamental
nature. 
In the case of SSSs, 
however,
the connection between the phenomenological definition and
the gross physical properties is not unique.
The broad range of temperatures and luminosities that characterize
SSSs can be associated with many different types of physical
systems. Analogous statements apply to the new class of QSSs. 
In this section we provide an overview of 
physical models for VSSs.

\subsection{WD Models for Supersoft X-Ray Sources}  

SSS-like emission is expected from hot WDs. Approximately half of the
SSSs with optical IDs
in the Milky Way and Magellanic Clouds are systems containing
hot white dwarfs: symbiotics, recent novae, the central stars
of planetary nebulae. In most of these systems, the high
temperatures and luminosities are associated with episodic
nuclear burning, or with cooling subsequent to
nuclear burning. The question 
that was raised by the discovery of binary SSSs like
CAL~83 and CAL~87 
is whether the soft luminous emission
from them and from some of the other mysterious SSSs is due to
 quasi-steady nuclear burning of matter accreted by a 
WD from a Roche-lobe-filling companion.

Conceptually, there is a natural place in the pantheon of accreting WDs, 
for WDs in contact systems with high-enough
accretion rates to permit 
quasi-steady nuclear burning.
Among the  accreting WD systems
 that dominated studies before the
discovery of SSSs
were cataclysmic variables (CVs) and symbiotics.
In CVs the accretion rates are low
(typically $\leq 10^{-9} M_\odot$ yr$^{-1}$) and 
the observed luminosities can be explained by accretion power.
The accretion rates are low because,
although the donor fills its Roche lobe,
 the mass ratio ($M_{donor}/M_{WD}$)
is typically small and the donor is typically not evolved.
In systems with larger mass ratios, the accretion rate should be 
larger, in some cases  large enough ($\sim 10^{-7} M_\odot$ yr$^{-1}$) 
that the accreting material would be burned
in a quasi-steady manner. Although the
appearance of such systems would be quite different from that
of normal CVs, especially because the luminosity
would be $10-100$ times larger than the accretion luminosity, which would itself be larger than typical, they could represent an
epoch of normal CV lifetimes, as follows.

\noindent (1) When Roche lobe
overflow of a main-sequence star more massive than the WD
begins, there  could be a brief ramp-up time
($\sim 10^5$ yrs), during
which the mass-transfer rate would be below the value needed 
for nuclear burning. The system might appear to
be a CV, although with a brighter donor star than
is typical of CVs. 

\noindent (2) The accretion rate reaches the value needed for
quasi-steady nuclear burning, driven by the thermal-time-scale
adjustment of the donor to a shrinking Roche lobe.
 There ensues an epoch during which high luminosities
would be generated by nuclear burning, with $k\, T$ in the SSS range;
the system would be
a close-binary SSS (CBSS). The CBSS era ends after $\sim 10^7$ years, 
roughly the thermal time scale of the donor, when the
mass ratio has reversed. After the system cools, the SSS
behavior ends. 

\noindent (3) The donor, which is now less massive
and which is no longer stressed
by a quickly shrinking Roche lobe,  
may underfill its Roche lobe for some time.
Eventually, either due to dissipative processes
or to its own evolution and expansion, the donor will again 
fill its Roche lobe and the system may again appear to be a CV. 

\noindent (4) If dissipative processes were
responsible for initiating the second epoch of mass transfer,
then the system would simply be a normal CV, following
a standard evolutionary track. 

\noindent We also note that

\noindent (5) If stellar evolution
played the key role in re-establishing contact, 
then the orbital period will increase as mass transfer proceeds.
in some systems, the donor's envelope would be exhausted
before the donor became very evolved. In others, the donor
would become a full-blown giant. In the latter case, 
some systems could again experience an epoch of high-mass transfer 
and SSS-like behavior. These systems would be wide-binary SSSs
(WBSSs). 

\noindent (6) Wide binary SSSs are more commonly expected when
the donor first fills its Roche lobe as a giant.
Some WBSSs would be virtually indistinguishable from
symbiotic systems.

The donor stars in most symbiotics are thought to feed the WD
through winds, rather than through Roche-lobe overflow. 
Because the donors are very evolved, the rate of wind capture
by the WD can be high enough that episodic or quasi-steady nuclear burning
can occur.
There is
good evidence that the high luminosities of symbiotics are
powered by nuclear burning, which may be episodic in some
systems or steady in others.

Thus, quasi-steady-nuclear burning WDs in which a companion
fills its Roche lobe is a natural extension of the class
of CVs, with CVs representing the lower-accretion-rate
systems. More massive and/or more evolved donors lead to
supersoft binaries. The Roche-lobe filling SSS binaries (SSBs) with the 
most evolved donors have much in common with symbiotic binaries.
SSBs therefore form a natural bridge between
CVs and symbiotics, as they comprise a natural extension 
of both classes, producing
a unified picture of accreting WDs.

\subsubsection{The Close-Binary SSS Model}

The close-binary SSS (CBSS) model was designed to explain the phenomenon of
binary SSSs which, like the flagship sources CAL 83 and CAL 87,
have orbital periods in the range of tens of hours, 
and estimated bolometric 
luminosities $\sim 10^{37}-10^{38}$ ergs s$^{-1}.$\footnote{
The range $\sim 10^{37}-10^{38}$ ergs s$^{-1}$ was suggested by early
analyses of the ROSAT data that used blackbody fits. Some of
these estimates were later revised when WD model atmosphere models were applied.}
 (See
van den Heuvel et al.\ 1992; Rappaport, Di~Stefano, \& Smith 1994 [RDS].) 
The range of observed orbital periods implies 
a range of Roche lobe radii compatible with the size and mass
of donors that could in fact meet these requirements.
The model is therefore roughly consistent with the data on the
known systems.    
At present, most of the known SSSs with optical IDs that do not
 place them in an already-understood-class of hot WD system are considered
to be candidates for the CBSS model.

First principles estimates based on a population synthesis study 
to determine how many such CBSSs should be active in a galaxy such
as our own found that there are likely to be on the order of
$1000$ presently active CBSSs in the Milky Way with 
$L> 10^{37}$ ergs s$^{-1}$ (RDS; 
Yungelson {\it et al.} 1996). Studies of the effect of
absorption then confirmed that all but a fraction of a percent of these
systems would not have been detected by ROSAT (Di~Stefano \& Rappaport 1994).  

In spite of the fact that the CBSS model is self-consistent, 
it has remarkably little direct observational support. 
There are, however, 
indirect signs that some CBSS candidates
may be well-described by the model. These signs include the following.

\noindent (1) There are well-defined regions in the H-R diagram where steady
nuclear burning has been predicted to occur. Determinations of
SSS temperatures and luminosities thus far have been subject to
significant uncertainties. Nevertheless,
some (but not all)
 CBSS candidates seem to occupy regions of the H-R diagram 
consistent with quasi-steady nuclear burning. 
At present, until Chandra's low-energy calibration is better 
understood and/or
more XMM data become available, the uncertainties are too large to
allow apparent placement in the H-R diagram to confirm or falsify
the conjecture that any particular CBSS candidate 
contains a nuclear-burning WD.   

\noindent(2) The X-ray spectra (most of which are still
fairly crude) are reasonably well-fit by WD atmosphere models
(van Teeseling \etal\ 1996). Both {\it Chandra} and {\sl XMM} 
grating spectra of selected SSSs are beginning to be analyzed,
and will allow us to test the applicability of
WD atmosphere models in detail. There is published work on
just one system, CAL 83 (Paerels et al.\ 2001).
There are clear disagreements between the WD atmosphere
models applied to the data so far and the observed grating
spectra. It remains to be seen if these can be resolved.                                           
           
\noindent(3) 
Nuclear-burning WDs should have accretion disks with
distinctive features (Popham \& \rd\ 1996). This is because
the amount of energy provided by the WD
in the form of heat and radiation,  
is $\sim 10$ times greater than the accretion energy.
The inner regions of such a disk are thick and very hot,
contributing to the total soft X-ray emission. The disk
geometry flares at large radii. It  cools with distance from the
central WD, and is the dominant  source
of radiation at ultraviolet and optical wavelengths. 
The optical and UV observations of several CBSS candidates
are consistent with the first-principles disk predictions,
and this may be the strongest indication that CBSSs
contain accreting objects surrounded by reprocessing-dominated
disks (Popham \& \rd\ 1996). 
In addition, the first-principles 
predictions of disk geometry are in general agreement
with a model of CAL 87 that was constructed to explain the 
eclipse profile of that system (Meyer-Hofmeister, Schandl, \& Meyer 1997).

\noindent (4) Some CBSS candidates have been observed to have jets
with velocities roughly compatible with the escape velocity from a WD.
Since, however, objects more compact than WDs
can be surrounded by WD-sized photospheres, it may be possible
for such winds to originate well below the photosphere, and to
escape from a region around the poles.
If so, the resulting velocities could be comparable to those
observed.

\subsection{Other Nuclear Burning WD Models}  

\subsubsection{Symbiotics}  
In symbiotic systems, the donor star is very evolved. 
In most cases it is thought
that the donor does  not fill its Roche lobe but that the WD accretes
mass by capturing a small fraction of the donor's wind. Since the
donor may be losing mass at $10^{-6}-10^{-4} M_\odot$ yr$^{-1}$,
the WD accretion rate can be high enough to allow either episodic
or quasi-steady nuclear burning. Symbiotics form a
well-studied class and there  is evidence from several
directions that the model described above is
correct. The existence of symbiotics is, therefore, one of the strongest
arguments that nuclear burning may fuel high luminosities and
temperatures comparable to those measured in SSS binaries.

\subsubsection{Nova-Like Variables and Bright CVs}
The effects of absorption make it difficult to discover SSSs in
the disk of the Milky Way; even bright, hot SSSs 
in the Galactic disk can be found in
surveys like {\sl ROSAT's All-Sky Survey} only if they are
 within about $1$ kpc of Earth (\rd\ \& Rappaport 1994).  
Fortunately, two methods that do not depend on X-ray surveys
have been able to identify previously unknown Galactic SSSs with
temperatures and luminosities lower than those of any discovered  
via surveys.  

\noindent{\it Vy Scl stars:\ } 
The pattern of optical time variability of the SSS RX J0513.9-6951
(Alcock et al.\ 1996, Southwell et al. 1996),
was observed by Brian Warner (private communication)
to be similar to that of members of
the 
Vy Scl subclass
of nova-like variables.
Because the soft X-ray emission from RX J0513.9-6951 
turned on during its optical low-state,
we monitored Vy Scl stars and obtained ToO time to
observe one such system, V751 Cyg, during its
optical low state. During this occasion, it was observed to
have a bolometric luminosity of 
$6.5 \times 10^{36}$ ergs s$^{-1}\, (D/500 pc),$ where $D$ is
the distance to the system, and $k\, T = 15^{+15}_{-10}$ eV
(Greiner et al.\ 1999, Greiner \& \rd)  
This luminosity is several orders of magnitude 
larger than expected due to accretion alone, making
a strong case for nuclear burning.

\noindent{\it Bright CVs:\ } 
If a CV appears so bright that its luminosity cannot
be explained by accretion alone, then it may be a candidate 
NBWD (Patterson et al.\, 1998,
Steiner \& Diaz 1998).
V Sge has been observed to exhibit a soft state
(Greiner \& van Teeseling 1998), but, 
for realistic values of absorption,
the estimated luminosity is far too low
($\sim 10^{32}-10^{33}$ erg s$^{-1}$) to describe a
canonical SSS.
Although by
invoking additional absorption, it is possible to find a physical
solution with $k\, T \sim 17$ eV and $L\sim 10^{36}$ erg s$^{-1}$,
the SSS nature remains unconfirmed.

\smallskip

Nova-like variables and other CVs are thought to
contain accreting WDs. The discovery of soft X-radiation
from V751 Cyg,  emitted at luminosities
far in excess of what what could be provided by accretion,
therefore constitutes indirect evidence that episodic or quasi-steady
nuclear burning may be responsible for the high luminosity
in some SSBs. Available evidence for V~Sge is somewhat weaker,
but may also be interpreted as supporting the SSB
model (van Teeseling \& Greiner 1998). 
There are a number of other CVs for which there is evidence of
energy emission from a CV in excess of the energy available
from accretion. Because nuclear burning near the
surfaces of lower-mass WDs should yield smaller
luminosities and lower temperatures, the WD mass in
SSSs in this category are likely to be smaller.
Note that V751 Cyg, and possible some of the 
other VY Scl stars, possibly V Sge and other high-L CVs,
appear to represent the low-L/low-T extension of
SSBs, and the high-L/high-T extension of CVs.  

\noindent
An interesting point is that, if a significant
fraction of nova-like variables or anomalously bright CVs   
exhibit SSS states, the total number of SSBs in the Galaxy
could be very large. For example, in a $1$ kpc region
surrounding the sun, these systems outnumber other CBSS candidates
by a factor of more than $10$. 
These systems could have been (indeed may already have been)
 detected in M31 (\rd\ et al. 2002a, 2003), but would not
have been detectable in the exposures taken of the 
$4$ galaxies we consider in the companion paper (\rd\ \& Kong 2003 b).

\smallskip

\subsubsection{Wide-Binary SSSs (WBSSs)}
The high accretion rates needed for episodic or quasi-steady
nuclear burning can also be achieved in contact binaries with
orbital periods larger than those of CBSSs. In these cases,
the nuclear evolution of a Roche-lobe filling giant
can drive the required high mass transfer rate (Hachisu, Kato,
\& Nomoto 1996; \rd\ \& Nelson 1996).    
While optical emission from CBSS binaries comes  predominantly
from the irradiated disk 
(Popham \& Di~Stefano 1996), SSSs with giant donors  
may be optically characterized by emission from the donor, and such
systems could therefore have detectable 
optical counterparts even in external galaxies.

\subsubsection{Irradiation-Driven Winds}  
Two SSBs, 1E0035.4-7230 and RX J0537.7-7034, have short orbital periods
($3-4$ hours) and mass functions that
consistent with a donor star of mass smaller than that of the WD.
In such systems, mass transfer could not be mediated by the
same processes as those assumed for a CBSS.
Instead, an irradiation-driven wind has been postulated as 
the source of mass donated to the WD 
(see, e.g., van Teeseling \& King 1998).  
Irradiation-driven winds are likely to play a role in most SSBs;
in the binaries described above, they are the drivers of mass transfer. 
 
\subsection{Alternatives to WDs}

\subsubsection{Neutron Stars}

Kylafis \& Xilouris (1993) found that near-Eddington accretion onto 
neutron stars could be associated with large photospheres,  
consistent with the effective temperatures of SSSs.
The extent of the radial flow can determine whether each
neutron star with accretion rate within $\sim 10\%$ of the Eddington
limit 
appears as an SSS (for an extended radial flow)
or as a standard X-ray binary dominated by harder emission.
Although it was not clear that the extended radial flow
and large photosphere were 
to be expected, 
neutron star models 
received observational support with the discovery of 
an unusual transient X-ray pulsar, RX J0059.2-7138 (Hughes 1994).
RX J0059.2-7138 
has been observed to have a soft ($\sim 30$ eV) unpulsed component that is
highly luminous ($\sim 6.7\times 10^{38}$ ergs s$^{-1}$; 
$N_H \sim 8.8 \times 10^{20}$ cm$^{-2}$. (See also 
Kohno, Yokogawa, \& Koyama 2000.)
In fact, if we were not in the pulsar's beam, RX J0059.2-7138 would
simply be classified as an SSS. 
It is unlikely that RX J0059.2-7138 is an anomaly. In fact,
Her X-1, a galactic X-ray X-ray pulsar, has long been
known to have a luminous time-variable soft component
(Shulman \etal\ 1975, Fritz \etal\ 1976). 

\subsubsection{Black Holes}

An accretion disk around an intermediate-mass BH is expected to yield 
SSS emission. 
Consider an accreting Schwarzschild BH of mass $M.$ 
If the accretion disk is optically thick but geometrically thin, 
and if the disk extends down to the last stable
orbit,
the temperature of its inner rim 
is roughly  
\begin{equation}  
%
k\, T = 42\, eV\,\Bigg({{1000\, M_\odot}\over{M}}\Bigg)^{{{1}\over{2}}}
    \Bigg[\Big({{0.1}\over{\alpha}}\Big)\, 
          \Big({{L_{obs}}\over{3\times 10^{37} erg s^{-1}}}\Big)\Bigg]
    ^{{1}\over{4}},   
\end{equation}  
where $M$ is the  mass of the BH, $\alpha$ is the efficiency factor,
and $L_{obs}$ is the observed luminosity. 
This equation illustrates that, for a broad range of values
of the BH mass and of the luminosity, SSS emission is expected.

It is less certain whether stellar-mass BHs can emit spectra
dominated by a soft component, but if extended photospheres are
possible for accreting NSs, they may also be possible for
accreting stellar-mass BHs. 
A BH nature has been suggested for
CAL 87, e.g., (Cowley, Schmidtke,
 Crampton, \& Hutchings 1990, 1998; Hutchings, Crampton, 
 Cowley, \& Schmidtke 1998)    
Orbital observations of CAL 87 have suggested that
the primary may be a star with mass between $1.3\, M_\odot$ and
$4\, M_\odot$. The mass estimate 
hinges on the assumed 
location from which line emission emanates. 
The highest estimated mass might correspond to a 
black hole (BH) accretor.

\subsubsection{Stripped Cores of Tidally Disrupted Stars}

At the center of most, if not all, galaxies is thought to
be a massive ($\sim 10^6-10^9 M_\odot$) BH.    
A star of mass $M_\ast$ and radius 
$R_\ast$ can be disrupted by approaching
within $R_t$ of the BH.
\begin{equation}
    R_t=
\Bigg( \eta^2 {M_{bh} \over {M_\ast}} \Bigg)^{{1}\over{3}}\, R_\ast,
\end{equation}
$M_{bh}$ is the mass of the BH, and
$\eta$ is a parameter of order of unity.
The disruption leaves an end-product,
the
giant's hot dense core.
The core remains hot ($T>10^5$ K) and bright ($L>10^2 L_\odot$)
for
$10^3-10^6$ years, thereby providing
the longest-lasting
signal of a tidal disruption (TD). 
This scenario was proposed by \rd , Greiner, Garcia,
\& Murray 2001. 
Given the expected rate of TDs, 
in a galaxy such as M31, several of these remnants
could be active at any given time.   
Some stripped cores are expected to be WDs or pre-WDs, and some 
are expected to be helium stars.

The possibility that SSSs in the center of nearby galaxies
could be signatures of TDs is interesting,
particularly because other signatures of TDs 
are so difficult to identify with confidence. The
complementary signature most considered is
an event due to the accretion of a portion
of the disrupted star's envelope by the BH
(Hills 1975, Lidskii \& Ozernoi 1979,
Gurzadyan \& Ozernoi 1980, Rees 1988). 
The associated accretion event
can last for months or decades, with luminosities possibly as
high as
$\sim 10^{44}-10^{46}$ erg/s.
There is a growing body of data on UV
and X-ray 
flares that may be consistent with these
sorts of accretion events (see references in \rd\ et al. 2001). 
It is nevertheless difficult to establish a definite link
between observed flare events and accretion events,
so information about stripped cores in nearby galaxies
would be important. 

The possibility of studying the stripped cores of
disrupted stars is an important motivation of the search
for SSSs in the central regions of galaxies.

\subsection{Models for Contaminants}     

In this section we have so far focused on models in which the
SSSs we discover are luminous X-ray binaries. We expect, however,
that other types of objects will produce the same broadband
X-ray signatures. SNRs form the primary class of SSS ``contaminants" that 
are luminous ($L_X > 10^{36}$ erg s$^{-1}$), and which are actually  
members of the galaxy being observed.
In M31, $2$ SSSs are SNRs. Interestingly, one of these M31 SNRs is
among the softest sources in M31. Most of the 
other $33$ M31 SSSs we   
identified using the criteria presented in this paper, are highly variable;
many are transients. We therefore know that SNR contanimants form
only a minor portion of the SSSs in M31. In \rd\ \& Kong (2003 a),
we studied the available data on the variability of SSSs in
$4$ more distant galaxies (M101, M83, M51, and NGC 4697; see also 
\rd\ \& Kong 2003 b). Although the limited time coverage of the 
observations we studied allowed only the brightest sources to be
checked for variability,
we did find evidence 
of variability on time scales
of  a year.  This is consistent with an X-ray binary nature
for the majority of bright SSSs.  

Knots in diffuse emission from the galaxy can also have very soft
spectra, and they may be misidentified as SSSs when they cannot be 
spatially resolved.
This is most likely to occur near the centers of galaxies with
a significant diffuse soft component, but can happen
in any location in which the X-ray emission appears to be dominated by
diffuse emission. If the sources are bright or the time sampling is good,
time variability can help to identify which sources in regions of diffuse 
emission may be X-ray binaries; observations at other wavelengths may
be helpful in finding counterparts to extended objects.
In the absence of such complementary information, however,
SSSs discovered in regions of diffuse emission should not be
assumed to be X-ray binaries.   

Other systems identified by our algorithm are
dim foreground objects or bright background objects.
Foreground stars can emit soft X-rays. In many cases such stars
will have been identified by optical surveys and can be
ruled out as luminous X-ray binaries. In high surface brightness
regions of the observed galaxy, however, the survey of foreground stars
may be less complete, and we may not be able to identify which SSSs
are foreground stars. A further complication is that
the soft X-ray emission from foreground stars can be
highly variable, so variability cannot be taken as
a signature that the SSS is an X-ray binary. Distant soft AGN can
also be selected as SSSs; observations at other wavelengths can 
help to identify some, but probably not all of these. Further,
some nearby magnetic CVs can also be selected as SSSs; it may
be difficult to identify such sources at other wavelengths.

The standard method to 
estimate the contribution of foreground and 
background sources is to use results derived from deep field surveys
(Giacconi et al.\, 2001, Brandt et al.\, 2001). Because, however,
 we are specifically interested in SSSs, which
have not yet been studied in the deep fields, 
we have used another approach, sketched below, and discussed in
more detail in (Di\, Stefano et al.\, 2003).
Briefly, we have applied our algorithm to {\it Chandra}
data from several fields analyzed by the 
ChAMP team. We consider only fields located
away from the Galactic plane, and containing no clusters or galaxies.
In such fields, we generally we find $1-3$ VSSs in the S3 CCD. 
When, therefore,
in observations of an external galaxy, we discover tens of VSSs in the S3 CCD,
we can assume that the majority of them are associated with the galaxy.  

Finally, we note that, although SNRs, foreground stars, and other
``contaminants" do not dominate the VSSs we identify with galaxies,
our algorithm does provide an efficient way to search for
X-ray active SNRs and for a subset of foreground stars.
   
\section{Phenomenological Definition of Luminous Supersoft X-Ray Sources}

Our phenomenological definition should select sources
described 
by the physical
models discussed above. The WD models alone
define a broad range of temperatures, from $k\, T < 10$ eV
up to $\sim 150$ eV.\footnote{The value $150$ eV is an absolute
maximum for nuclear-burning WD models. It corresponds to the Eddington luminosity
emitted from an effective radius equal to that of a Chandrasekhar mass WD.
In fact, the photospheric radius is expected to be larger; only
if some of the radiation is scattered to higher energies, or if X-rays are emitted by a small portion of the photosphere,
would such
a high effective temperature be possible.}
 For example, while V751 Cyg had a best fit    
temperature with $k\, T < 10$ eV, a $1.4\, M_\odot$ WD
with Eddington-luminosity nuclear burning on its
surface would have $k\, T \sim 150$ eV.
NBWD luminosities range from $\sim 10^{35}$ erg s$^{-1}$ up
to the Eddington limit for a $1.4\, M_\odot$ object 
($\sim 2 \times 10^{38}$ erg  s$^{-1}$). The stripped core
of a high-mass star might have a temperature near the low end
of the temperature range, but a luminosity in excess of $10^{39}$
erg s$^{-1}.$ Accreting BHs could have even higher
luminosities, with temperatures in the SSS range or even higher.      

\subsection{Unique Features of SSSs}     

The sensitivities of the detectors used for X-ray astronomy
tend to peak for photons with energies near or above $1$ keV.
Until the advent of {\it Einstein} and then {\it ROSAT},
it was difficult to study sources with energy distributions
peaked significantly below $1$ keV. 
It was also difficult to detect and study such sources by 
using detectors effective at UV or EUV wavelengths, 
since 
this radiation is readily absorbed by the ISM.
When, therefore, the {\it ROSAT}  
{\it All Sky Survey}  
discovered several dozen sources with little or no emission above
$1$ keV, it was an important result.

Figure 1 illustrates the key features of SSS spectra, versus the
spectra of other X-ray sources.
We consider a sequence of thermal models, with temperatures 
ranging from $k\, T = 20$ eV to $k\, T = 2$ keV within each 
large panel,
and $N_H$ ranging from $4.0 \times 10^{20}$ cm$^{-2}$ to
$2.5 \times 10^{22}$ cm$^{-2}$ from the top left panel to the
bottom right panel. Spectra of sources with $k\, T < 100$ eV are shown
in red, as these would generally be considered to be SSSs;
green marks sources with $100$ eV $ < k\, T < 1$ keV, while blue
shows the spectra for the hottest sources we considered.
Each model was used to simulate a source located in M31 (at a distance
of $780$ kpc). We used PIMMS  
to compute the expected count rate. The variable along the 
vertical axis is the log of the number of counts detected by ACIS-S in 
$10$ ksec. Values along the
horizontal axis correspond to bin numbers: $1 = 0.3-0.5$ keV,
 $2 = 0.5-0.7$ keV,  
 $3 = 0.7-0.9$ keV,  
 $4 = 0.9-1.1$ keV,  
 $5 = 1.1-1.5$ keV,  
 $6 = 1.5-2.0$ keV,  
 $7 = 2.0-7.0$ keV.  

For small values of $N_H,$ represented here
by $N_H=4.0 \times 10^{20}$ cm$^{-2}$, the spectra of SSSs
peaks in bin 1 (photon energies $< 0.5$ keV) and declines to
$0$ by bin $6$ (photon energies between $1.5$ and $2.0$ keV).  
The decline is steepest for SSSs of the lowest temperatures;
$25$ eV sources cut off above $0.5$ keV, 
$40$ eV sources cut off above $0.7$ keV, 
$55$ eV sources cut off above $0.9$ keV, 
and so on. For $100$ eV sources, there are a small number of
counts above between $1.5$ and $2.0$ keV. The comparison with
sources of higher temperature ($k\, T > 250$ eV) is striking: 
the detectable spectrum of high-temperature sources
 is either peaking or still rising
in bins where the SSSs are cut off. As $N_H$ rises, 
photons are lost in all bins, but the erosion is sharpest at
lowest energies. The total count rate for SSSs declines sharply
with increasing $N_H$, with contributions from the lowest
bins falling to $0$ first.  

The spectral characteristics of a $100$ eV sources are significantly
different from those of a $250$ eV source. 
If we are to use
a phenomenological  
definition of SSSs, it is important to identify a temperature 
between these two temperatures  
that marks the transition between SSSs with
thermal spectra and  other thermal sources. 
Figure 3 shows results parallel to those of Figure 2,
but for $k\, T$ between $125$ eV and $225$ eV. At $175$ eV,
there is some hard emission in bin 7 for all values of $N_H$.
We have therefore chosen  $175$ eV to define the boundary
temperature for SSSs: sources with $k\, T < 175$ eV are SSS.
It is likely that a thermal spectrum does not provide a good
fit to all SSSs. We therefore also consider power-law models.
Because sources with $\alpha \geq\ 3.5$ are rare, we choose this 
as our definition of power-law-fit SSSs.

\subsection{A Phenomenological Definition of SSSs}  

If any source has enough counts that a spectrum can be well fit, then
the source is an SSS if any of the following conditions are met.

\noindent (1) The spectrum is well-fit by a 
blackbody model with $k\, T < 175$ eV, 

\noindent (2) The spectrum is well-fit by a 
power-law model with $\alpha \geq 3.5$, 

\noindent (3) Whatever the best fit model, 
less than $10\%$ of the energy
is carried by photons with energy greater than $1.5$ keV.

The reason for adding condition (3) is that, even if the
emitter in an SSS can be well-described by either a 
thermal or power-law model,
interactions with matter in the vicinity of the emitter
can upscatter some of the photons. The choice of $10\%$
is rather arbitrary; future studies of larger numbers of
SSSs may suggest that this criterion be altered. For now
we allow as much as $10\%$ of the energy to come in at energies
above $1.5$ keV, because this is roughly
consistent with efficient reprocessing or with, e.g.,
 the
presence of a companion emitting a strong shocked wind.
We do not allow higher values, simply because we want to
reserve the SSS classification for  sources whose spectrum is clearly
dominated by the soft component.

Most sources in external galaxies provide too few photons for a
meaningful spectral fit. In these cases, we must try to
select sources that would satisfy the above criteria, had we been able to 
collect enough
photons.
The selection criteria are described in section $4$ 
and in the flowchart of Figure 3.   
We propose that any source that passes the test 
defined by these procedures
be provisionally identified as a very soft X-ray source.

\section{The Selection of Supersoft Sources} 

\subsection{Overview of the Selection Criteria}

In this section we outline our selection criteria. Because we define  9
sub-categories, the results may seem complicated. We therefore begin by  
pointing out that the categories have a simple overall structure.
Details are provided in the Appendix.

\subsubsection{``Classical'' Supersoft Sources} 

There are $2$ separate set of criteria that 
select sources most likely to be as soft as the standard SSSs first
discovered in the Magellanic Clouds. These $2$ sets of  conditions
are the HR conditions and the $3\sigma$ conditions.
In a separate paper 
we have written about the HR conditions and their results when applied to $4$
galaxies:  M101, M83, M51, and NGC4472.
The $3\sigma$ conditions mirror the HR conditions,
but should be able to identify SSSs with slightly lower count rates. Any
source satisfying either the HR and $3\sigma$ conditions  will be
referred to as a ``classical'' SSS or simply as an SSS.

\subsubsection{Quasisoft Sources} 

Many X-ray sources which are very
soft may not satisfy either the HR or $3\sigma$ conditions. Consider,
e.g. a 125eV source located in M31. If it is luminous enough, we will
detect a significant number of photons in the M band. If, in addition, the
source is located behind a large gas column, we may receive few photons in
the S-band. In addition, because {\it ROSAT} did not have good high-energy
sensitivity ($> 2$ keV) and because we do not know enough about physical
models for SSSs to eliminate the possibility that some have a
relatively hard tail, it is important to retain and classify very soft
sources that do not satisfy either the HR or $3\sigma$ conditions. We
therefore introduce 2 categories of weaker conditions. Any source
satisfying one of the weaker conditions
is referred to as a quasisoft source (QSS).
Each QSS sub-category is defined by specific conditions described
in the appendix. A brief explanation of each category is given here.

There are two types of QSS selection criteria. The first
requires that there be little or no hard emission ($H/{\Delta\, H} < 0.5$)
Any source fulfilling this condition falls into one of the so-called ``noh" 
categories.  
Depending on how dominant the S and M bands are,
QSSs with little or no hard emission
 sources are labeled NOH (for no hard emission), MNOH (no hard emission
but a $3 \sigma$  detection in M, even though the broadband spectrum is
dominated by soft photons), SNOH (no hard emission, little M emission, but
a $3 \sigma$ detection in S), and FNOH 
(the spectrum is
flatter than in the other noh categories.)   

The second type of QSS selection criteria identifies sources
which may emit some hard radiation ($H/{\Delta\, H} > 0.5$),
but in which the soft component clearly dominates. These
sources are labeled HR$_1$, $3\sigma_1$, or $\sigma$.

\subsection{Distinctions between SSS and QSS}

It is important to note that, in many cases, the only distinction
between SSSs and QSSs may be operational - i.e. in the way members of each
category are selected. There are some cases in which identical sources
observed under different condition, may be put in different (QSS vs SSS)
categories. As {\it Chandra's} low-energy sensitivity declines, e.g., many
sources that would have qualified as SSSs if observed in AO1 will be
selected as QSSs, even if their luminosity and spectral characteristics
are constant. Other QSSs may have spectra that are genuinely harder than
SSSs - they may, e.g., have higher temperatures. ($kT\sim 200$ eV) or
they may be dominated by very soft radiation, but may included a
power-law tail.

It is a separate question, however, whether QSSs and SSSs have different
physical natures. Any SSSs which are nuclear burning WDs should not have
$kT \gaeq 150-175$ eV. Hotter sources are therefore unlikely to be
nuclear-burning WDs. Accreting NSs or accreting intermediate-mass BHs
could, however, have spectra that put them in either the SSS or QSS category.

\subsection{Tests} 

There are two questions we must address. First,
do the conditions described above select a large fraction of
genuine SSSs? Second, what fraction of and what 
kinds of non-SSS sources are
mis-identified as SSSs?  
 
To answer these questions, we have tested the selection criteria on
a set of thermal models. We consider sources
with intrinsic luminosities between $6 \times 10^{35}$ ergs s$^{-1}$
and $1 \times 10^{38}$ ergs s$^{-1},$ computing their
fluxes as if they were in M31 (i.e., $\sim 780$ kpc from us),
computing the number of counts in each of the 
$7$ bins described in \S 3.1, plus one bin $0.2-0.3$ keV. 
We considered a 35 ks observation, and added two $\sigma$
uncertainties. We considered $10$ luminosities, and for each 
luminosity, we considered $10$ realizations, randomizing the
number of counts in each bin ($0-7$)
between $-2\, \sigma$ and $2 \sigma$ of the 
PIMMS-computed value. We used the PIMMS software to 
compute the counts that would be detected, using $4$
values of $N_H$ for each source: 
$4.0 \times 10^{20}$ cm$^{-2}$ (red points), 
$1.6 \times 10^{21}$ (yellow; open circles) cm$^{-2}$,
$6.4 \times 10^{21}$ cm$^{-2}$ (green; larger open circles),  
$2.5 \times 10^{22}$ cm$^{-2}$(blue; largest open circles). 
The lowest value is somewhat smaller
than typical Galactic values of $N_H$ along directions to M31.  
 The largest value would correspond to directions for which
the absorption within M31 was exceptionally high.

The numerical results are summarized in Tables 1, 
and in Figure 4. Figure 4 clearly shows that
the conditions HR, $3\, \sigma$, NOH, MNOH, and SNOH
produce few misidentifications for thermal sources. That is,
almost all selected sources with more than $10$ counts and without
a large intervening absorbing column ($6.4 \times 10^{21}$ cm$^{-2}$
or $2.5 \times 10^{22}$ cm$^{-2}$) have $k\, T < 175$ eV. Some of
the higher temperature SSSs ($k\, T$ near or above $100$ eV)
are missed, however, particularly when $N_H$ is large.
The additional conditions, HR$_1$,
$3\, \sigma_1,$ $\sigma,$ and FNOH help to identify these high-T,
high-$N_H$ SSSs. These further conditions, however,
yield a higher rate of identifications of
systems with $k\, T$ on the order
of a few hundred eV. Some of these medium-temperature QSSs,   
may be SNRs, but those we have been able to study in M31 appear to
be variable on time scales of months to years, ruling out
SNR models (Di\,Stefano et al. 2003b).
Overall, $90\%$
of SSSs (blackbodies with $k\, T < 175$ eV)
are correctly identified; only $70$ of $2000$``hard" 
($k\, T  > 500$ eV) were misidentified; $75\%$ of QSSs 
($ 175 \leq\ k\, T < 500$ eV) were also selected.

\subsection{{\sl XMM} Observations}

The above selection criteria were designed to optimize
the selection of SSSs as simulated by PIMMS for {\it Chandra}
ACIS-S observations. As presently formulated, they will have 
almost the same effect on data from {\sl XMM}, 
which offers similar energy
coverage but with larger effective area.  
We compared 
the sensitivity of {\sl
XMM}/PN with a thin filter and {\it Chandra} ACIS-S, we found that
for SSSs, in
the S, M, and H bands, 
{\sl XMM} collects $\sim 3$, $\sim 2$, and 
$\sim 3$ times as many photons as
{\it Chandra}, respectively.
This suggests that the primary difference between 
{\sl XMM}/PN and {\it Chandra} ACIS-S is the former's somewhat
smaller relative sensitivity to photons in the $M$ band. Thus, 
fraction of sources in the SSS-FNOH
and SSS-MNOH categories may be smaller when {\sl XMM}
rather than {\it Chandra} data is used.
Because of its larger effective area,
{\sl XMM} may do better in searching for SSSs 
in regions where source confusion and diffuse gas emission
are not problems.

\subsection{Real {\it Chandra} Observations}

As noted in \S 5.2, {\it Chandra's} ACIS
detectors have suffered degradation at low ($<$ 1 keV) energies.
This degradation is time dependent and was not included in the
PIMMS simulations. In the companion paper, we apply the 
selection algorithm to real {\it Chandra} ACIS-S data.
Their effect will not be the same as the effect they had
on the simulated data. For example, when the counts in the $S$ bin
are half of what they would have been in the simulated data,
while $M$ and $H$ are relatively unaffected, imposing
the requirement that
$S > 3\, M$, will be the same as if we had imposed 
$S > 6\, M$ in the simulations. That is, the low-energy degradation
makes it more difficult for an SSS to be identified.
In many cases, a source that, without the degradation,
might have been identified at a higher point in the hierarchy
(e.g., as an SSS-HR source), will now be identified 
at a lower point in the hierarchy (perhaps as an SSS-MNOH or
SSS-$\sigma$ sources). In other cases, SSSs will simply fail to 
be identified as such. There is no direct cure
for this latter problem, because we cannot assume that
a lack of low-energy photons is due to a lack of sensitivity.
Fortunately, even with the degradation, we are able to identify 
a significant population of SSSs in each galaxy.  

\section{Summary and Conclusion}
The primary focus of the research described in this paper is to
develop a systematic procedure
to identify SSSs in external galaxies. To accomplish this
we began by crafting an operational definition of SSSs.
(See \S 3.2.) It is based entirely on the spectral shape:
a blackbody spectral fit with $k\, T < 175$ eV,
or a power law spectral fit with $\alpha > 3.5$, or
any spectral fit where less than $10\%$ of the energy is carried by
photons with energy $> 1.5$ keV.
These criteria can be applied directly to sources providing enough counts to
allow a good spectral fit. Since most sources in distant galaxies do
not provide enough counts, we have designed a set of conditions
 which
can
be applied algorithmically to
determine if the true spectrum of a low-count source is likely to
satisfy the SSS criteria.
Because the conditions work for high-count sources as well
as low-count rate sources,  we simply apply our algorithm to
all sources in each galaxy to identify the SSSs.
The sources are then sub-classified according to the
specific conditions they satisfy.

The strongest conditions are called the HR and $3\sigma$ conditions.
We have called a source an SSS if it satisfies
 either the HR or $3\sigma$ conditions.

We call all sources that satisfy the
weaker conditions, QSSs. Some QSSs may be similar to SSSs, but perhaps more highly absorbed,
or they may be genuinely harder than the canonical SSSs.
The harder sources may have values of $k\, T$ between roughly $175$ eV and $350$ eV,
or may be very soft sources which include a low-luminosity hard
spectral component.
The QSSs with dominant spectra that are genuinely harder than those
of SSSs (e.g., those with $k\, T > 175$ eV) are almost certainly not WDs,
unless the X-ray emission emanates from only a small portion
of the photosphere. They may be accreting NSs or, perhaps
most likely, accreting intermediate-mass BHs.

Tests with simulated data show that our criteria 
accomplish the following.
(1) They select a majority of SSSs--
$90\%$ of thermal models.
(2) They select only a small fraction of
 ``hard'' sources ($4\%$ 
of models with $k\, T > 500$ eV).
(3) They select a substantial fraction of quasi-soft sources.
For blackbody models, quasi-soft sources were those in
the range $175$ eV $< k\, T < 500$ eV; $70\%$ of these
were selected.

The selection of QSSs is difficult to
      avoid if we are to succeed at identifying
possibly absorbed  high
temperature SSSs ($k\, T$ near $100$ eV), some of which
could be high-mass accreting WDs on their way to becoming
Type Ia supernovae.
On the other hand, the presence of QSSs in the data
is intriguing, because there appear to be no known X-ray binary analogs
with similar luminosities ($> 10^{36}$ erg s$^{-1}$) in the Galaxy
or Magellanic Clouds.  
Some may be SNRs; repeated X-ray observations and
optical emission line studies can distinguish SNRs
from X-ray binaries. Those that vary in time are likely to be X-ray binaries
either of a type or in a state that is not typical of nearby X-ray binaries.
The discovery of such sources in real data sets thereby
challenges us to understand what they are. 

In the companion paper we describe how we tested the selection algorithm by applying it to
real data from $4$ galaxies. We verified that the spectra of the
sources identified as SSSs tend to be in the range
we used to define SSSs. These investigations
have also yielded interesting information about extragalactic SSSs,
providing some first steps toward answering the $7$ questions
posed in \S 1.3.

\begin{acknowledgements}
We are grateful to Pauline Barmby, Mike Garcia,
Jochen Greiner, Roy Kilgard, Miriam Krauss, Jeff McClintock, Koji Mukai,  
Paul Plucinsky, Will Pence, Andrea Prestwich, Frank Primini, Roberto Soria, Douglas Swartz, Harvey Tananbaum, Ben Williams, Kinwah Wu,
and Andreas Zezas 
for stimulating discussions and comments. 
We are grateful to Paul Green and the ChaMP collaboration
for providing
blank field source data. 
This research has made use of the
electronic catalog of supersoft X-ray 
sources available at URL http://www.aip.de/~jcg/sss/ssscat.html
and maintained by J. Greiner.    
RD would like to thank the Aspen
Center for Physics for providing a stimulating environment,
and the participants of the 2002 workshop on  
{\it Compact Object Populations in External Galaxies}  
 for insightful comments.  
This work was supported by NASA under AR1-2005B, GO1-2022X,
and an LTSA grant, NAG5-10705.  
\end{acknowledgements}

\appendix{}

\medskip
\centerline{\it Selection Criteria} 
\medskip

The selection criteria described below are also illustrated in
the flow chart of Figure 3.
 
Three energy bins are used to define the hardness ratios: 
{\bf S:} 0.1-1.1 keV, {\bf M:} 1.1-2 keV, {\bf H:} 2-7 keV. We employ $2$
hardness ratios.  
\begin{equation} 
HR1={{M-S}\over{M+S}}   
\end{equation} 
\begin{equation} 
HR2={{H-S}\over{H+S}}   
\end{equation} 

The selection criteria will sometimes use the numbers of counts, $C(i),$
 in
each of the $7$ bins defined in \S 3.1. In addition, we consider 
bin $0: 0.1-0.3$ keV. Although photons with energies in this lowest bin
cannot be used to compute spectral characteristics, especially
without a well-understood low-energy calibration, they can be
useful to the identification of SSSs.

\subsubsection{Hardness Ratio Conditions}
 
The strongest set of conditions we impose are strict
hardness ratio (HR) conditions. We demand that $HR1 < -0.8$
and $HR2 < -0.8.$ These conditions imply that  
\begin{equation}
S > 9\, M,  
\end{equation}
\begin{equation}
S > 9\, H.  
\end{equation}
We also require
\begin{equation}
HR1_{\Delta} < -0.8, 
\end{equation}
\begin{equation}
HR2_{\Delta} < -0.8, 
\end{equation}
where 
$HR1_{\Delta}=
[(M+\Delta M)-(S+\Delta S)]/[(M+\Delta M)+(S+\Delta S)],$ 
and $\Delta S$ and $\Delta M$ are the one-$\sigma$ uncertainties in
$S$ and $M,$ respectively.  An analogous expression defines 
$HR2_{\Delta}.$ The conditions on $HR1_{\Delta}$ and $HR2_{\Delta}$
can be satisfied only by systems that have small relative 
uncertainties in the total numbers of counts. This generally
implies either a high count rate or else a total absence of flux
in the $M$ and $H$ bands, combined with low background in the
$S$ band. 
We denote systems that satisfy the HR conditions ``SSS-HR."

When tested on thermal models as described in \S 4.3,
the HR conditions select only systems with $k\, T < 175$ eV.
SSS-HR sources are therefore prime SSSs.

\subsubsection{$3\sigma$ Conditions}

To accommodate SSSs with lower count rates, we implement the
following ``$3\sigma$" conditions. First, we demand
a two $\sigma$ detection is the $S$ band.
\begin{equation}
{{S}\over{\Delta\, S}} > 2  
\end{equation}
In addition,
\begin{equation}
{{S}\over{\Delta\, S}} > 3\, {{M}\over{\Delta\, M}}  
\end{equation}
\begin{equation}
{{S}\over{\Delta\, S}} > 3\, {{H}\over{\Delta\, H}}  
\end{equation}

When the uncertainties in $S,$ $M,$ and $H$ can be
approximated as $\sqrt{S},$ $\sqrt{M},$ and $\sqrt{H},$
respectively, then the $2$ conditions above imply the
same restrictions on the relative values of $S$ and $M$ and $S$
and $H$ as do Eqns.\, 3 and 4. 
We denote systems that satisfy the $3\sigma$ conditions 
``SSS-$3\sigma$."

The HR and $3\sigma$ conditions select sources based on the
dominance of flux in the S band. 
If, however, there is a large column of gas
between the source and the detector, the preferential absorption
of soft  photons erodes the signal in $S$ relative to that in the
other bands, so that even genuine SSSs may satisfy neither condition.
Especially if the flux is small, the HR and $3\sigma$ conditions
will fail to identify many SSSs.
We therefore introduce weaker conditions that can identify
additional SSSs, but which may also identify some harder sources.
Because the SSS status of sources identified via weaker conditions is
less secure, we refer to all such sources as ``quasisoft" sources (QSSs).
Any QSSs with high enough count rates to allow spectral fits, and for
which the fits shows a low effective temperature, can then be
reclassified as SSSs.

To proceed, we turn to another hallmark of SSSs:
the lack of
high-energy emission from many of the ``classical" SSSs.

We use the condition,
\begin{equation} 
{{H}\over{\Delta H}} < 0.5.    
\end{equation}
to identify $2$ branches in the identification procedure.
If the condition is satisfied,    
we know that the source cannot be hard, and devise further tests 
to determine if it is soft. We refer to the path defined by 
these further tests
as {\it Branch 1}. 
If the condition given in Eq.\, (10) is not satisfied, it is still possible for the 
source to be an SSS. {\it Branch 2} implements those conditions that can select
SSSs while rejecting hard sources.

\subsubsection{Branch 1: NOH}  
 
If $C(6)/\Delta\, C(6) \leq 1$ and
$C(5)/\Delta\, C(5) \leq 2,$ then there
is at most a weak detection in the bins
corresponding to medium energies.
Sources satisfying these conditions are
called ``QSS-NOH''; they exhibit little or no
emission above $1.1$ keV. 

If the NOH condition is not satisfied, then there is
a detection above  $1.1$ keV. 
The condition $M/\Delta\, M > 3,$ creates another
bifurcation which we can follow along $2$
sub-branches, {\it Branch 1.1} and {\it Branch 1.2}.

\centerline{\it Branch 1.1: MNOH}

If we have reached this point along the path, there is no hard emission,
but there is significant emission in the $M$ band. 
We can call the source 
``supersoft" only if emission in the $S$ band
dominates and if, there is a steep decline
in the number of counts from the low to high
energy bins. We therefore require that 
${{S}\over{\Delta\, S}} > 2\, {{M}\over{\Delta\, M}}$ and
$C(5) > 2.6\, C(6).$ Sources satisfying these conditions 
are designated ``QSS-MNOH".

\centerline{\it Branch 1.1: FNOH}
Sources
that do not satisfy the conditions for QSS-MNOH could be SSSs
which are highly absorbed, 
or they could be sources with
spectra that can best be viewed as flat in M relative to S --i.e., 
little emission in the soft or hard bands, e.g., a 
$200$ eV blackbody. We call such sources ``QSS-FNOH".

\centerline{\it Branch 1.2: SNOH}

If 
we have reached this point along the path, there is no hard emission,
but there is also little flux in the $M$ band. The source
can be SSS only if there is emission in the  
$S$ band. If $S/\Delta\, S > 3,$ we use the designation ``QSS-SNOH".

\centerline{\it Branch 1.2: FNOH}

If the source fails the SNOH test, it is placed in the QSS-FNOH category.

\subsubsection{Branch 2: HR$_1$} 

In this case, $H/\Delta\, H > 0.5,$ so there may be some hard emission.
If the hard emission is nevertheless a small component of the spectrum,
the source could still be an SSS. First we require that 
either (a) $H/T < 0.005,$
where $T= S + M + H,$ or (b) both $S/\Delta\, S > 3$ and $H/\Delta\, H < 1$.    
In addition, we require the same conditions on HR2 required for the 
``HR conditions": $HR2 < -0.8$ and $HR2_{\Delta} < -0.8.$ At this point,
however, we relax the conditions on HR1. That is, we allow for the
possibility that absorption has eroded the flux in the S band relative
to the flux in the M band. The new conditions on HR1 are: 
$HR1 < -0.5$ and $HR1_{\Delta} < 0.$ These conditions could, e.g., be satisfied by
an absorbed $100$ eV SSS. Sources satisfying these conditions will be referred to
as ``QSS-HR$_1$''.
      
\centerline{\it Branch 2: ``$3\, \sigma_1$'' conditions} 
The $3\, \sigma_1$ conditions represent a somewhat relaxed version
of the $3\, \sigma$ conditions. We require a two-$\sigma$ detection
in $S$, $S/\Delta\, S > 2,$ and that the hard flux be a small fraction
of the total flux: $H/T < 0.005.$ As before, we require
$S/\Delta\, S > 3\, H/\Delta\, H.$  But we relax the condition (8), 
replacing it with
$S/\Delta\, S > M/\Delta\, M.$  Systems satisfying this set of conditions
are designated ``QSS-$3 \sigma_1$".

\centerline{\it Branch 2: ``$\sigma$" conditions}
Sources with $H/T < 0.05,$ and $S\over \Delta\, S$ $> 3$ are
designated ``QSS-$\sigma$''.

\centerline{\it Branch 2: NOT SSS} 
The remaining sources are, for the purposes of this classification
scheme, not SSSs.

\begin{deluxetable}{lcccc}
\tablewidth{0pt}
\tablecaption{Test of the selection algorithm}
\tablehead{\colhead{Thermal} & \colhead{All} & \colhead{kT $<$ 175 eV}
& \colhead{175 eV $\leq$ kT $<$ 500 eV} &\colhead{kT $\geq$ 500 eV}
}
\startdata
 $N_{total}$    &      6800  &      3200   &     1600   &     2000\\
 Low L  &       918  &       828   &       49  &        41\\
 HR      &      1329  &      1328   &        1  &         0\\
 $3\sigma$ &       143  &       136   &        7  &         0\\
 NOH     &       128  &       116   &        7  &         5\\
 SNOH    &        53  &        50   &        3  &         0\\
 FNOH     &       459  &       230   &      210  &        19\\
 MNOH    &        97  &        62   &       35  &         0\\
 HR$_1$  &       299  &        75   &      224  &         0\\ 
 $3\, \sigma_1$&       248  &        42   &     206   &        0\\
 $\sigma$&       520  &        16   &      500  &         4\\
 Not SSS \& QSS  &      2605  &       317   &      358  &      1930\\   
\enddata
\tablecomments{ 
$N_{total}$ is the total number of sources generated. 
``Low L'' sources yielded $4$ or fewer counts; we did not 
consider these sources further. 
One thermal model with more than $4$ counts
exhibited flux in only the hard band ($S+M=0;$) we did not 
consider these sources further.}
 
\end{deluxetable}

\begin{figure}
{\rotatebox{-90}{\psfig{file=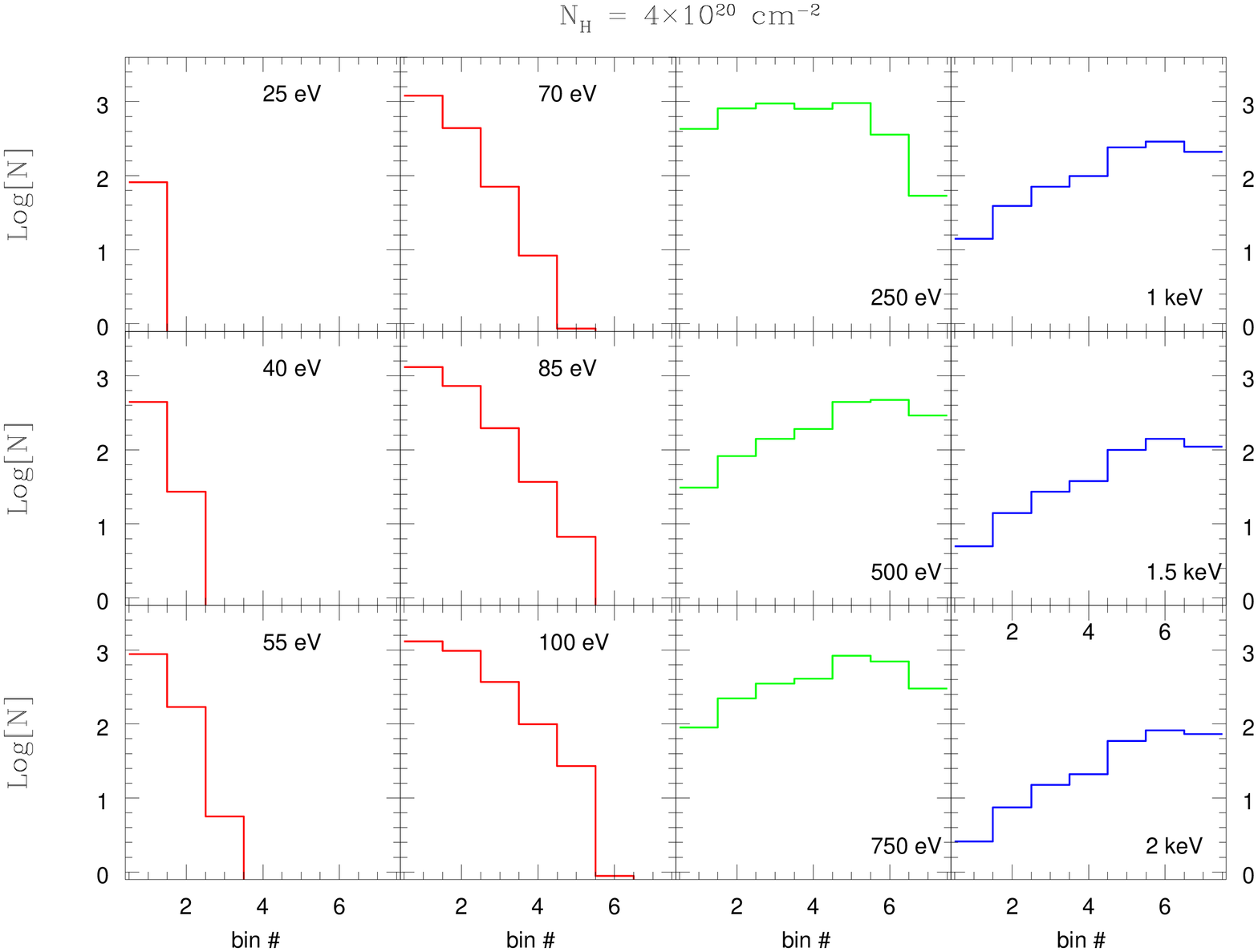,width=3.3in,height=3.7in}}}
{\rotatebox{-90}{\psfig{file=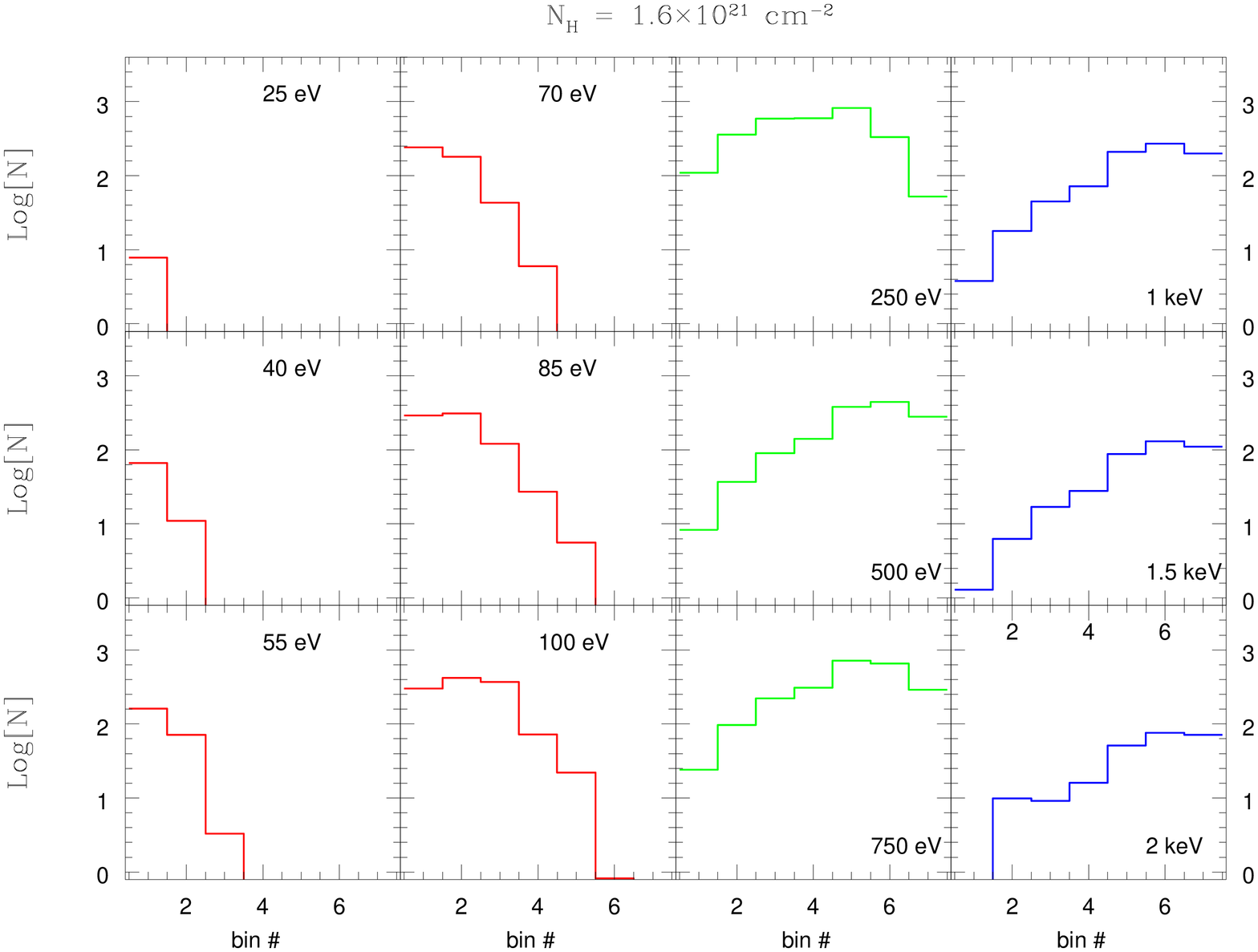,width=3.3in,height=3.7in}}}
{\rotatebox{-90}{\psfig{file=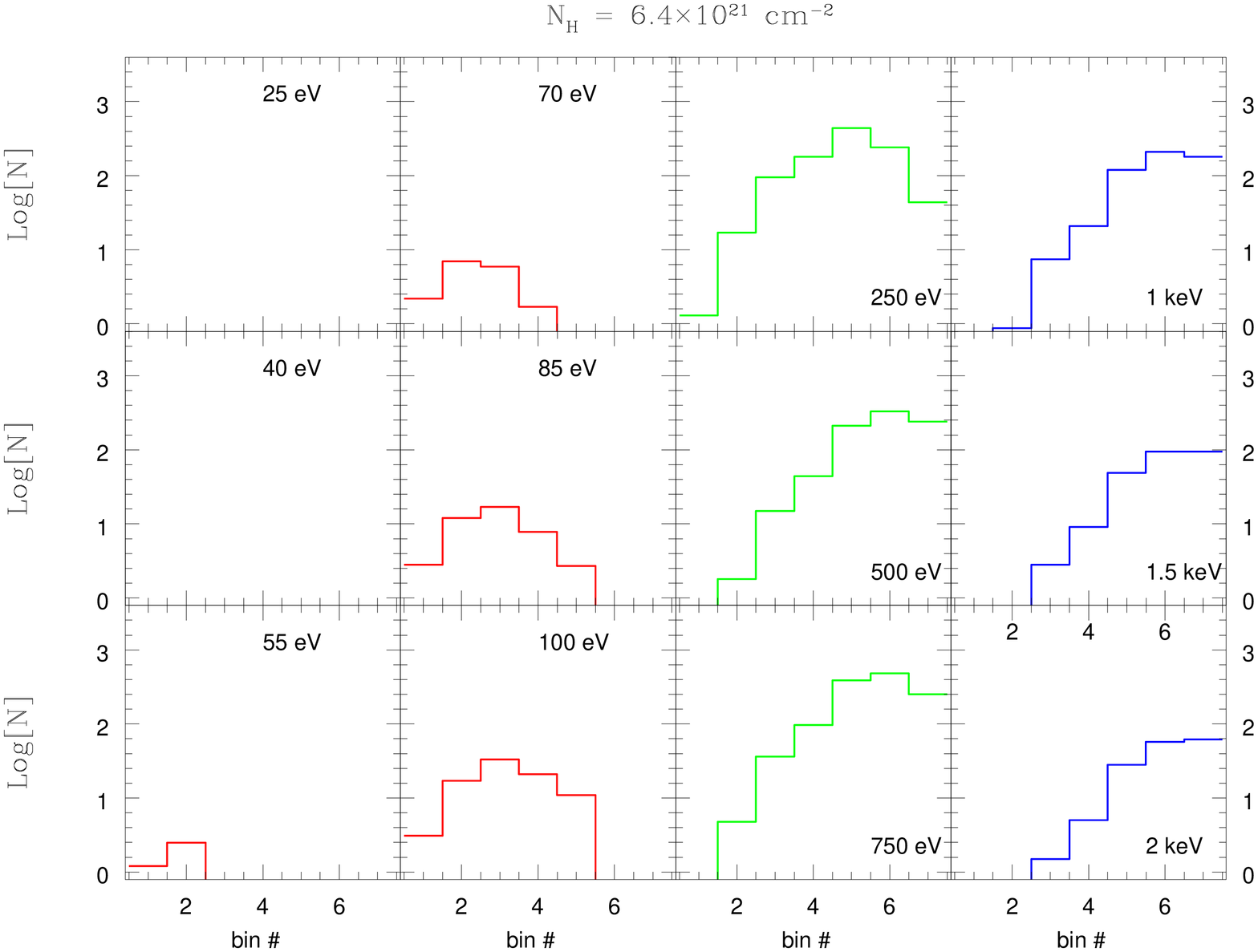,width=3.3in,height=3.7in}}}
{\rotatebox{-90}{\psfig{file=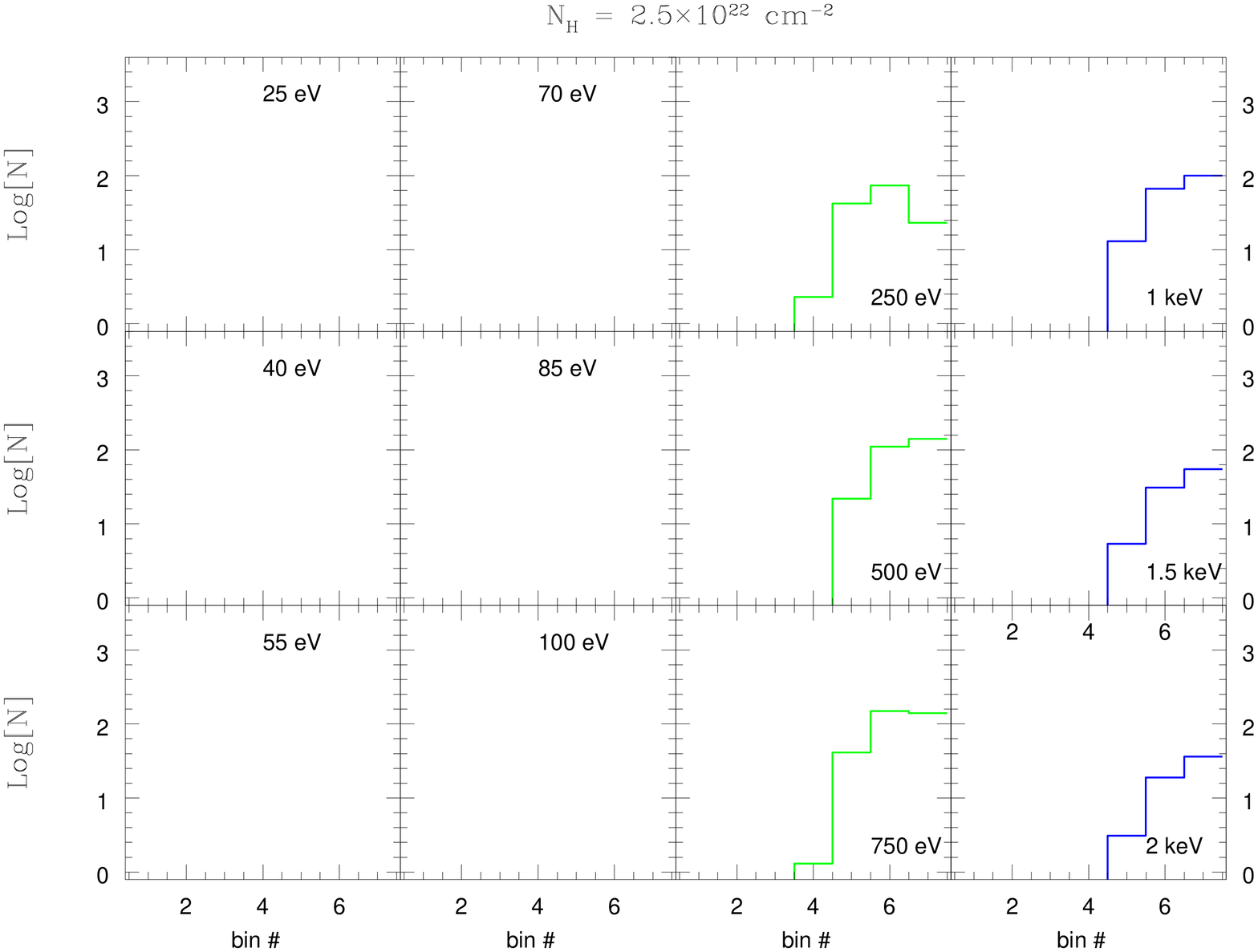,width=3.3in,height=3.7in}}}
\caption{PIMMS Results: $Log[N]$ vs energy bin number. $N$ is the
number of counts detected in each bin when a $10^{38}$ erg s$^{-1}$ 
source is placed in M31 ($d=780$ kpc)
and is observed for $10$ ksec. The energy bins are defined in \S 3.1.}
\end{figure}

\begin{figure}
\psfig{file=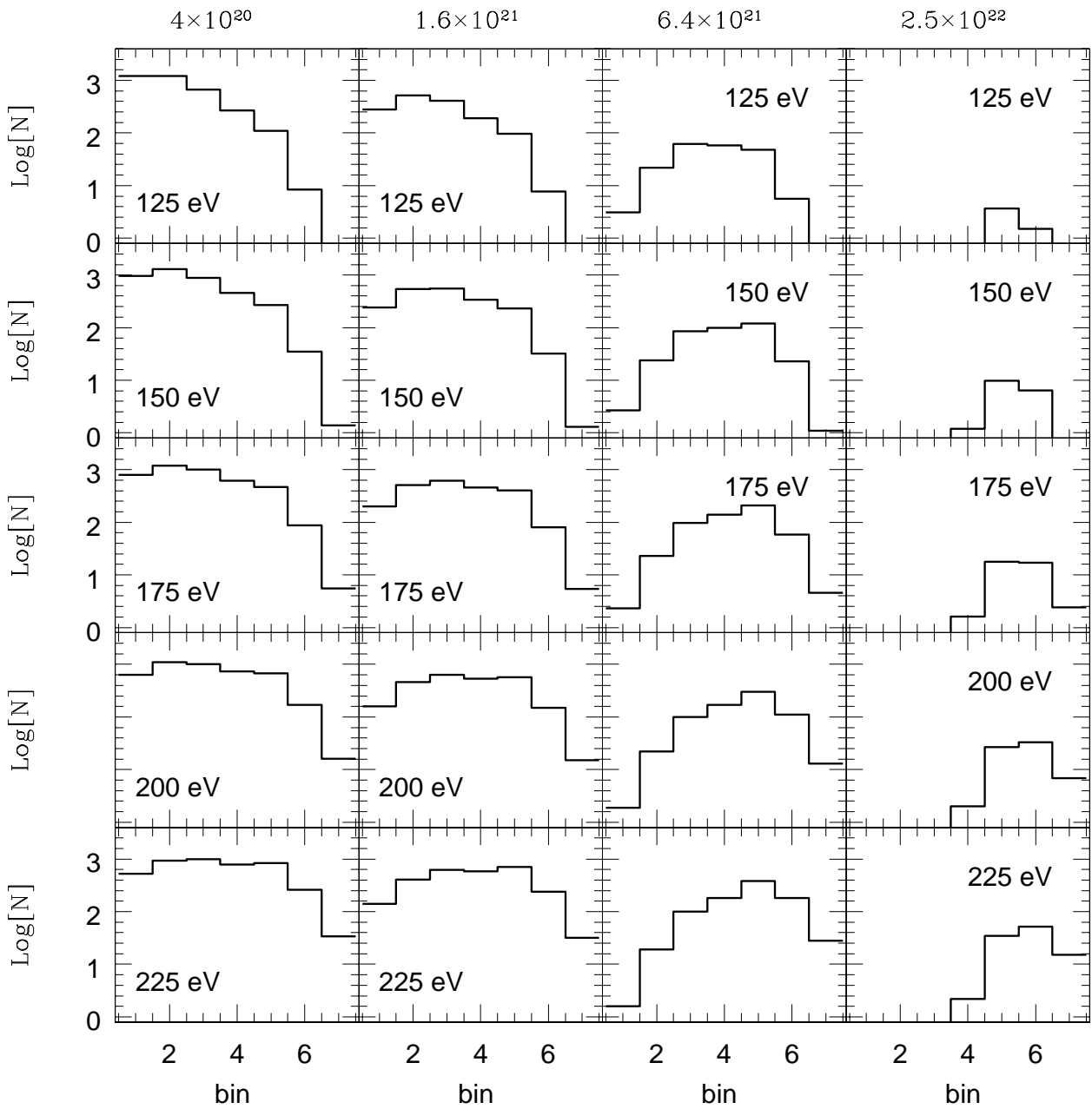,height=7in}
\caption{Same as Figure 1.}
\end{figure}

\begin{figure}
\psfig{file=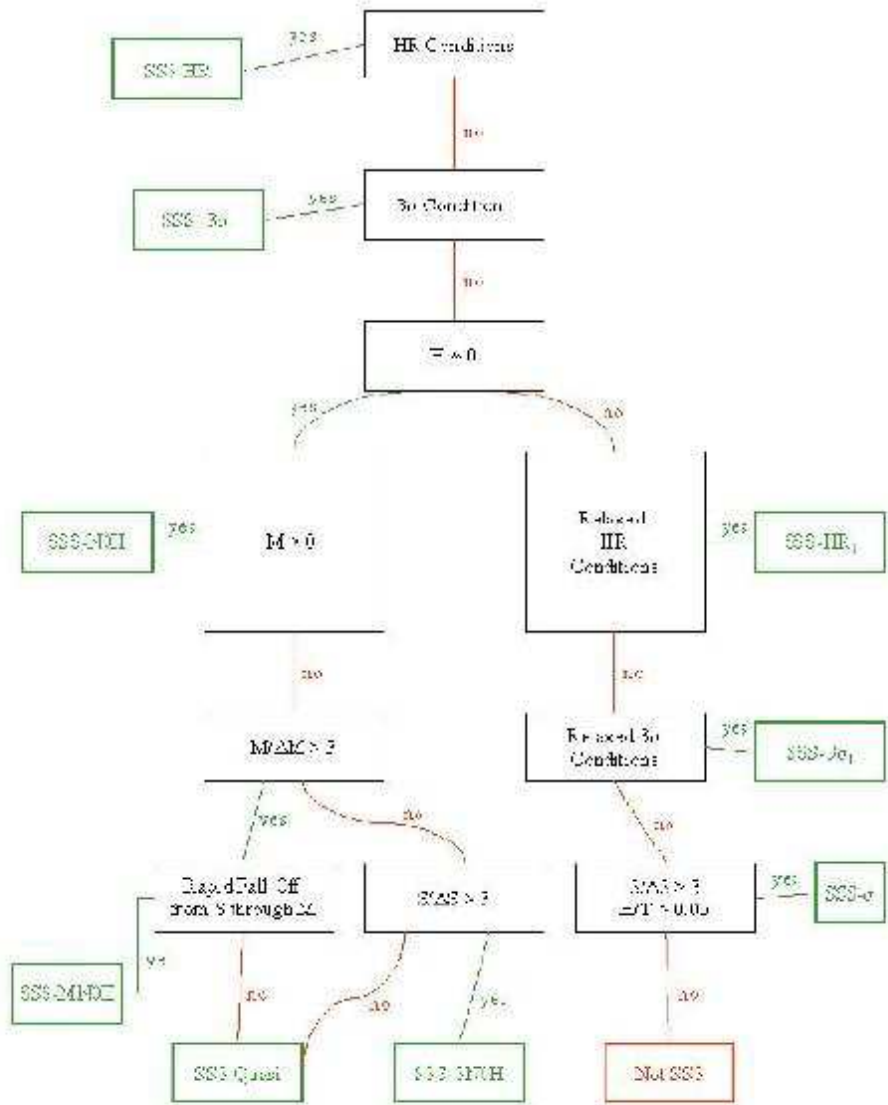,height=8in}
\caption{Flow chart of the selection algorithm. In principle, 
spectral fits of all high count rate sources can be used
to identify the brightest SSSs even before the HR conditions 
are applied.  In practice, we have found that the high count rate sources
were identified by the selection process sketched here.}
\end{figure}

\begin{figure}
\psfig{file=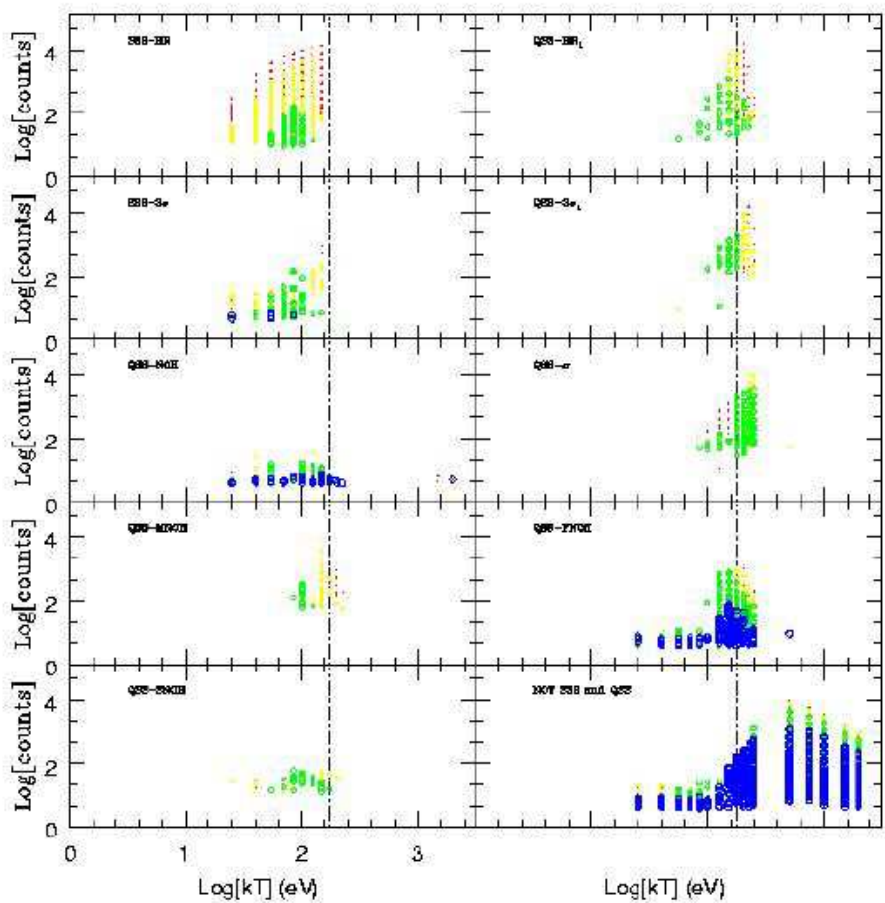,height=7in}
\caption{Test of the selection criteria as applied to
thermal models. Along the vertical
axis is the number of counts that PIMMS predicts would be detected
for M31 sources ($D=780$ kpc) with luminosities ranging from
$10^{36}$ ergs s$^{-1}$ to $10^{38}$ ergs s$^{-1}$. 
As described in \S 4, $4$ values of $N_H$ were considered:
$4.0 \times 10^{20}$ cm$^{-2}$ (red points),
$1.6 \times 10^{21}$ (yellow; open circles) cm$^{-2}$,
$6.4 \times 10^{21}$ cm$^{-2}$ (green; larger open circles),
$2.5 \times 10^{22}$ cm$^{-2}$(blue; largest open circles).
The logarithm of the temperature is
plotted along the horizontal axis. Points are shown only for 
those sources that would be identified as SSSs by the condition
labeled in the box; the lower left panel shows sources that
are not selected as SSSs and QSSs by any set of conditions.
}
\end{figure}


\end{document}